\documentclass[twocolumn,twoside,slac_two]{revtex4}
\usepackage{graphicx}
\usepackage{fancyhdr}
\renewcommand{\headrulewidth}{0pt}
\renewcommand{\footrulewidth}{0pt}

\newcommand{\eV}{\mbox{$\rm eV$}}

\newcommand{\nueb}{\mbox{$\bar{\nu}_{e}$}}

\newcommand {\be}{\begin{equation}}
\newcommand {\ee}{\end{equation}}
\newcommand {\bea}{\begin{eqnarray}}
\newcommand {\eea}{\end{eqnarray}}
\newcommand{\bdec}{\mbox{$\beta$ decay}}
\newcommand{\nnbb}{\mbox{$0\nu \beta \beta$}}
\newcommand{\tnbb}{\mbox{$2\nu \beta \beta$}}

\newcommand{\mtwo}{\mbox{$m_\nu^2$}}

\newcommand{\mnue}{\mbox{$m(\nu_e)$}}

\newcommand{\mee}{\mbox{$m_{\rm ee}$}}
\newcommand{\me}{\mbox{$m_{\rm e}$}}
\newcommand{\Mnue}{\mbox{$M_{\nu}$}}

\newcommand{\Thalfnonue}{\mbox{$T^{0\nu}_{\rm 1/2}$}}
\newcommand{\Gnonue}{\mbox{$G^{0\nu}$}}
\newcommand{\Mnonue}{\mbox{$M^{0\nu}$}}

\newcommand{\etal}{\mbox{\it et al.}}

\newcommand{\PRL}{\mbox{Phys. Rev. Lett.}}
\newcommand{\PL}{\mbox{Phys. Lett.}}

\newcommand{\ket}[1]{|#1\rangle}

\begin{document}

\title{Experiments for the absolute neutrino mass measurement}

%

\author{Markus Steidl}
\affiliation{Forschungszentrum Karlsruhe, Germany\\
Postfach 3640, 76021 Karlsruhe\\
eMail: markus.steidl@ik.fzk.de}

\begin{abstract}
Experimental results and perspectives of different methods to measure the absolute mass scale of neutrinos are briefly reviewed.
The mass sensitivities from cosmological observations, double beta decay searches and single beta decay spectroscopy differ in sensitivity and model dependance.
Next generation  experiments in the three fields reach the sensitivity for the lightest mass eigenstate of $m_1<0.2 \,\mbox{eV}$, which will finally answer the question if neutrino mass eigenstates are degenerate. This sensitivity is also reached by the only model-independent approach of single beta decay (KATRIN experiment). For higher sensitivities on cost of model-dependance the \nnbb \, searches and cosmological observation have to be applied. Here, in the next decade sensitivities are approached with the potential to test inverted hierarchy models.
\end{abstract}

\maketitle

\thispagestyle{fancy}


\section{\label{intro}Introduction}
The absolute neutrino mass is directly connected to  important questions in particle physics and cosmology.\\
In the established theory of neutrino mixing three different mass eigenstate $\ket{m_i}$ exist, which compose the three leptonic eigenstates $\ket{\nu_\ell}$. The knowledge of the absolute mass of at least one mass eigenstate or one neutrino flavour opens new windows for progress in the mentioned fields:  For verification of mass generating models of the Standard Model it would be a crucial test if the models predict correctly the still unmeasured hierarchy or degeneracy amongst the mass eigenstates. Within the hierarchical models, the classification into normal hierarchy and inverted hierarchy  is another benchmark for verification between such models. Also cosmological important questions like the composition of the energy density of the universe, or the theory of large scale structure evolution rely on the knowledge of the absolute neutrino masses.\\

Here in this article, current and scheduled experiments, which aim to measure the neutrino mass with sub-eV sensitivity are listed and compared against each other. An introductionary section of neutrino mixing (sec.\ref{sec_mixing}) is given to emphasize that the masses derived from cosmology (sec.\ref{sec_cosmo}), neutrinoless double beta decay searches (sec.\ref{sec_nnbb}) and single beta decay spectroscopy (sec.\ref{sec_singleb}) differ in their meanings. In sec.\ref{sec comparison} the systematic differences amongst the methods are compared.

\section{\label{sec_mixing} Neutrino masses and mixing}
The question if neutrinos are even massive has been answered clearly positive by neutrino oscillation experiments in recent years. These results establish, that the leptonic neutrino eigenstates $ $ are superpositions of 3 mass eigenstates $\ket{m_i}$:
$$
\ket{\nu_\ell} = \sum_i U_{\ell i}\,\ket{m_i} \label{mixing}
$$
The mixing matrix U can be parameterized as a product of a 3x3 matrix (with 4 mixing angles and 1 CP violating phase) and a diagonal matrix $(e^{i\alpha_1 /2},e^{i\alpha_2/2},1)$ with 2 Majorana phases $\alpha_{1,2}$\cite{pdg}.
Neutrino oscillations experiments are exclusively sensitive to the 3x3 matrix and the differences of the squared mass eigenstates $\Delta m_{ij}^2=m_j^2-m_i^2$.\\
 By measuring the mass of one neutrino flavor (e.g. $\ket{\nu_e}$) the masses of the other two flavors can be derived through the mixing matrix and the $\Delta m_{ij}^2$ as measured or constrained from oscillation experiments. The laboratory measurements focus exclusively on the electron neutrino mass $m_{\nu_e}$ as here the highest experimental sensitivity on the mass is given.

The existence of neutrino mixing means  that care has to be taken in comparing "neutrino masses" from different observations. The observable in single beta decay experiments (see sec.\ref{sec_singleb}) $\me^2$ is given as
\begin{equation}\label{me}
 \me^2 = \sum _i |U^2_{ei}|^2 m_i^2
\end{equation}
The masses derived from neutrinoless double beta decay searches (\nnbb):
 \begin{equation}\label{mee}
 \mee^2 = |\sum _i U^2_{ei} m_i|^2
\end{equation}
instead are dependant on the Majorana phases $\alpha_{1,2}$.  Only in the case of vanishing Majorana phases, \mee \, equals \me, otherwise $\mee <\me$. The neutrino mass \Mnue \ derived from cosmological observation
 \begin{equation}\label{Mnue}
 \Mnue = \sum _i m_i
\end{equation} is the sum over the mass eigenstates independent of the mixing matrix U.\\

From the neutrino oscillation measurements boundaries on \me,\mee \, and \Mnue \, can be inferred due to the constraint $\Delta m_{32}^2+\Delta m_{21}^2+\Delta m_{13}^2=0$ \cite{pdg}. This leads to the following benchmarks for experimental sensitivities:
\begin{itemize}
  \item If \me (\mee,\Mnue) \, is below 0.2 (0.1,0.6) \eV , degenerate models are excluded. Furthermore, neutrino mass does not play a significant role in structure formation and the contribution to the energy density is negligible ($<0.7\%$).
  \item Experiments with sensitivities of  $\me(\mee,\Mnue)< 50 (20,100)$\, m\eV \ have the potential to exclude inverted mass models.
  \item A lower bound of  $\me\geq 5$ m\eV can be inferred, equivalent to $\Sigma m_\nu\ge 50$ m\eV, independant of mass hierachy. For \mee \, the lower bound can approach zero due to the mentioned cancellation effects by the Majorana phases.
\end{itemize}

\section{\label{sec_cosmo} Limits and Sensitivity from cosmology}
\subsection{Method}
The neutrino density $\Omega_{\nu}$ is one parameter out of 11 in the standard cosmological model\footnote{the exact number of parameters defining the cosmological model differs from author to author}. The density is related to the number of massive
neutrinos and the neutrino mass by
$$\Omega_\nu h^2 = \frac{\Sigma m_\nu}{93.2 eV} \label{omeganu}  $$
where h is the Hubble parameter in units of 100 km/s/Mpc. As expressed by eq.\ref{omeganu}  cosmological data determines the incoherent sum of all neutrino mass eigenstates.
Massive neutrinos contribute to the cosmological matter density $\Omega_m$, but get non-relativistic
so late that perturbations in neutrinos up to scales around the causal horizon at
matter-radiation equality is suppressed.
This neutrino free streaming
leads to a suppression of mass fluctuations on small scales relative to large.
Thus, to extract the cosmological observable $\Sigma m_\nu$ any measurements of spatial matter distributions respectively its power spectrum are a sensitive tool.
Nevertheless, degeneracies of $\Sigma m_\nu$ with other parameters exist, which can be broken or constrained by inclusion of additional cosmological data.

The galaxy-galaxy power spectrum from Large Scale Structure (LSS) surveys is by now the most often used measurement to access matter distributions. At present there exist two large galaxy surveys with the Sloan Digital Sky Survey (SDSS)\cite{sdss,sdss2} and the 2 degree Field Galaxy Redshift Survey (2DFRG)\cite{twodf}. The statistics and systematic understanding of these data samples allowed the first observation of the Baryonic Acoustic Oscillation (BAO)peak \cite{baosdss,bao2dfrg}- well known from CMB observations- in galaxy distributions. Including BAO helps to break the degenarcy of $M_\nu$ with the number of neutrino species $N_\nu$ but also determines$\Omega_m$  more reliable \cite{raffelt06}. Power spectra of matter fluctuations on smaller scales can be inferred from Lyman $\alpha$-forest (LYA) data - the absorption observed in quasar
spectra by neutral hydrogen in the intergalactic medium.
Currently the most precise measurement of the LYA power spectrum comes
from the Sloan Digital Sky Survey [19, 20].

For breaking degeneracies amongst the cosmological parameters, the LSS data is preferably combined with data from the Cosmic Microwave Background (CMB). Here, use of WMAP data is standard but also further inclusion of other data sets (e.g. ACBAR,CBI,VSA,BOOMERANG), which are more sensitive to high multipole modes, exist.
Additionally some authors include Supernova 1a data. The SNLS catalog is widely used in this context.

\subsection{Status}
Table 1 shows examples of different analyses. The table claims certainly no completeness of the many analysdes published in the recent years, especially since the release of WMAP, 2dFGRS and SDSS data. Nevertheless, it shows the diversity of extracted mass limits on $\Sigma m_\nu$. Upper limits $(95\% \mbox{C.L})$ in the range of $[1,2] \eV$ can be inferred when analyzing exclusively single data sets or combining them just with one other. When combining several data sets sub-eV limits are obtained, where those in the range $[0.6,0.9]\eV$ are regarded as robust and are often quoted in publications. By adding more information the limits can even be pushed. Nevertheless, this enhances model dependance. For example the author of reference \cite{seljak2006} yields as upper limit $\Sigma m_\nu<0.17{\rm eV}\,(95\% \mbox{C.L})$ by combining LSS and CMB data with LYA data. On the other hand reference \cite{allen2003} even find a $2\sigma$-effect for non-zero neutrino mass by combining LSS and CMB data with x-ray data from galaxy clusters.\\
Cosmological approaches show a high sensitivity on the neutrino mass $\Sigma m_\nu$. The results are model-dependant  not only in the context of the underlying cosmological model but also of the used data sets to fix the multi parameter space of the underlying cosmological model.

%

\begin{table*}[t]
\begin{center}
\caption{Limits on neutrino masses from cosmology (top panel), \nnbb (middle panel), and single beta decay (bottom panel). The first column describes the experimental approach, the second column the specific experiment. The third columns gives the number of used parameters in the data fit in the case of cosmology respectively the exposures in the case of \nnbb. The fourth column gives derived limits on neutrino masses with references quoted in fifth column .}
\begin{tabular}{|l|c|c|c|c|}
\hline
\multicolumn{5}{|c|}{\textbf{Cosmology}}\\
\hline
 \textbf{Observation} & \textbf{Data sets}& \textbf{No. of} & \textbf{$M_\nu$} &\textbf{Ref.}  \\
                             &                   &   \textbf{Parameters}         & $(95\% \mbox{C.L})$&  \\
                             \hline

\hline \tiny{LSS} &  \tiny{2dFGRS} & 5 & $<$1.8 eV  & \cite{elgaroy2002} \\
\hline LSS,CMB  &  \tiny{2dFGRS, WMAP(1y),ACBAR,VSA,CBI} & 7 & $<$1.2 eV  & \cite{sanchez2006}  \\
\hline LSS,CMB  &  \tiny{SDSS, WMAP(3y)} & 9 & $<$0.9 eV  & \cite{tegmark2006}  \\
\hline LSS,CMB,SN1a, BAO & \tiny{2dFGRS,SDSS,WMAP(3y),SNLS,BOOM}  & 11 & $<$0.62 eV & \cite{raffelt06} \\
\hline LSS,CMB,SN1a & \tiny{2dFGRS,SDSS,SNLS } & 7& $<$0.66 eV & \cite{spergel2007}  \\
\hline LSS,CMB,SN1a, BAO,Lya, & \tiny{2dF,SDSS, SDSS(gal),SNLS, WMAP(3y), CBI,VSA,ACBAR} & 7 & $<$0.17 eV  &  \cite{seljak2006} \\
 \hline LSS,CMB,x-ray cluster data &  \tiny{2dFGRS,WMAP(1y),ACBAR,CBI,Chandra} & 10 & $=0.56^{+.30}_{-.26}$ eV  & \cite{allen2003} \\
\hline
\multicolumn{5}{}{}\\
\hline
\multicolumn{5}{|c|}{\textbf{\nnbb}}\\ \hline
Isotope                   & Experiment        & Exposure        &        \mee   $(90\% \mbox{C.L})$        &              \\  \hline
$^{76}{\rm Ge}$,enriched  & IGEX              & 8.9 kg y        &        $<[0.33,1.35]$ \eV & \cite{igex} \\
$^{76}{\rm Ge}$ enriched  & Heidelberg-Moscow & 36 kg y         &        $<[0.32,1.00]$ \eV & \cite{klapdor1} \\
$^{76}{\rm Ge}$ enriched  & Heidelberg-Moscow & 72 kg y         &        $=0.32\pm 0.03$  \eV & \cite{klapdor2}\\
$^{130}{\rm Te}$          & Cuoricino         & 3.1 kg y        &        $<[0.2,0.7]$ \eV & \cite{cuore} \\
$^{100}{\rm Mo}$          & Nemo-3            & $\sim 7.5$ kg y & $<[0.7,2.8]$   \eV  & \cite{nemo}\\
\hline
\multicolumn{5}{}{}\\
\hline
\multicolumn{5}{|c|}{\textbf{Single Beta Decay}}\\ \hline
Isotope                   & Experiment        &         &        $m_\beta$   $(95\% \mbox{C.L})$  &              \\  \hline
$^{3}{\rm H}$, solid state & Mainz,  MAC-E filter             &         &       $<2.3$ \eV                        &     \cite{mainz}       \\  \hline
$^{3}{\rm H}$, gaseous     & Troitsk,  MAC-E filter         &         &       $<2.1$ \eV                        &     \cite{troitsk}     \\ \hline
$^{187}{\rm Re}$, solid     & Mibeta, cryogenic detector            &         &        $<15$ \eV                        &     \cite{mibeta}      \\ \hline
$^{187}{\rm Re}$, solid     & MANU, cryogenic detector              &         &        $<26$ \eV                        &     \cite{manu}      \\ \hline
\multicolumn{5}{}{}\\
\hline
\end{tabular}
\label{example_table_2col}
\end{center}
\end{table*}

\begin{table*}[t]
\begin{center}
\caption{Perspectives on neutrino masses from cosmology (top panel), \nnbb (middle panel), and single beta decay (bottom panel). In case of cosmology representative examples are given. The quoted experiments are under construction or within a R$\&$D phase. If ranges are given in brackets, these are due to uncertainties of nuclear matrix elements.}
\begin{tabular}{|l|c|c|c|c|}
\hline
\multicolumn{5}{|c|}{\textbf{FUTURE Cosmology}}\\ \hline
Observation                      &     Data set              &             &        \Mnue   $(90\% \mbox{C.L})$        &              \\  \hline
CMB, LSS                         & Planck+ todays LSS data   &             &        $<200$ m\eV & \cite{plancknow} \\
CMB, Shear surveys               & Planck+ LSST              &             &        $<100$ m\eV & \cite{planckshear} \\
CMB, LSS         s               & Planck+ future LSS data   &             &        $<50-100$ m\eV & \cite{plancklss} \\
\hline
\multicolumn{5}{}{}\\
\hline
\multicolumn{5}{|c|}{\textbf{FUTURE \nnbb}}\\ \hline
Isotope                          &     Experiment        & Mass              & \mee   $(90\% \mbox{C.L})$&              \\  \hline
$^{76}{\rm Ge}$,enriched         & GERDA, Phase 2 of 3   & 0.1 t             & $<[90,290]$ m\eV          & \cite{gerda} \\
$^{76}{\rm Ge}$ enriched         & Majorana, demonstrator& (0.03-0.06) t     & $<100$ m\eV               & \cite{majorana2} \\
$^{150}{\rm Nd},^{82}{\rm Se}$   & Super-Nemo            &  0.1-0.2 {\rm t}  & $<[50,100]$ m\eV          & \cite{supernemo}\\
$^{130}{\rm Te}$                 & Cuore                 & 0.75 t            & $<30$ m\eV                & \cite{cuore2} \\
$^{100}{\rm Mo}$                 & MOON                  & 0.12 t            & $<70$    m\eV             & \cite{moon}\\
$^{136}{\rm Xe}$, liquid         & EXO200                & 0.2 t             & $<[133,186]$   m\eV       & \cite{exo200}\\
$^{48}{\rm Ca}$                  & Candles III           & 0.3 t of CaF$_2$  & $<500$   m\eV             & \cite{candles}\\ \hline
\multicolumn{5}{}{}\\
\hline
\multicolumn{5}{|c|}{\textbf{FUTURE Single Beta Decay}}\\ \hline
Isotope                          &     Experiment        & Inventory        &        \mee   $(90\% \mbox{C.L})$        &              \\  \hline
$^{3}{\rm H}$, gaseous           & KATRIN                & 24 g &        $<200$ m\eV & \cite{katrin} \\ 
$^{187}{\rm Re}$, solid            & MARE II             &  $~200$ g   &        $<90$ m\eV & \cite{mare22} \\ \hline
\end{tabular}
\label{example_table_3col}
\end{center}
\end{table*}
\subsection{Perspectives}
Weak lensing effects open an additional window to reconstruct the mass power spectrum. Hereby, it has to be distinguished between the weak lensing of CMB photons being scattered on the gravitational wells of the matter distribution and the weak lensing of photons emitted from galaxies. The former one leads to a subtle smearing of the CMB peaks at high multipoles $(l>1200)$, the latter one to a distortion of the visible galaxy shapes (shear effects).
The Planck satellite is the next scheduled CMB survey (launch in 2009) with full sky coverage and improved sensitivity to high multipoles and thus with sensitivity to weak lensing effects. In reference \cite{plancknow} a $2\sigma$ detection threshold of $\Sigma m_\nu<0.2\,{\rm eV}\,(95\% \mbox{C.L})$ is simulated when combining Planck data with the actual LSS data. By combining Planck data with Shear surveys, e.g. from LSST \cite{lsst} scheduled to operate in 2015,  the sensitivity can be pushed down to $\Sigma m_\nu<0.10 \,{\rm eV}\,(95\% \mbox{C.L})$ according to ref. \cite{planckshear}. A similar sensitivity of $\Sigma m_\nu\leq (0.05-0.1)\,{\rm eV}\,(95\% \mbox{C.L})$ \cite{plancklss} is expected by combination of Planck data with LSS data of next generation surveys focusing on high redshifts .\\
Thus, these analyses than explore a mass range, where a positive signal for $\Sigma m_\nu$ is expected independent of an inverted or non-inverted mass hierarchy in the neutrino sector.

\section{\label{sec_nnbb} Limits and Sensitivity from neutrinoless double beta decay}
\subsection{Method}
Double beta decay is an allowed rare transition between two nuclei with the same mass
number (A) that changes the nuclear charge (Z) by two units. The decay
only occurs if the initial nucleus is less bound than the final one, and both must
be more bound than the intermediate nucleus. These conditions are fulfilled in
nature for many even-even nuclei, and the double beta decay has been observed for many isotopes.

On the other hand, the neutrinoless decay,
\begin{equation}\label{2nue}
 (Z,A) \rightarrow (Z-2; A) + 2e^-
\end{equation}
violates lepton number conservation and is therefore forbidden in the standard
electroweak theory.
The process is
mediated by an exchange of a light neutrino, which must be a Majorana particle.
The experimental signature of the process is the simultaneous emission of  2 electrons, where the sum of their kinetic energies add up to a monoenergetic line at the position of the Q-value of the decay. Thus, the experimental observable are number or upper limits of signal counts or equivalent half-lives $\mbox{T}_{1/2}$.
The decay rate is
proportional to the square of the effective Majorana mass \mee :
\begin{equation}\label{0nue}
 \Thalfnonue^{-1} = \Gnonue \cdot |\Mnonue |^2 \cdot  \mee ^2
\end{equation}
\Gnonue denotes the exact calculable  phase space factor and \Mnonue \ is the matrix element for the nuclear transition, which must be theoretically calculated, as they are not related $1:1$ to the measurable matrix elements in normal double beta decay. The Majorana mass \mee \ is given by:
\begin{equation}
 \mee = |\sum _i U^2_{ei} m_i|^2
\end{equation}
Thus, in \nnbb experiments cancellations due to complex phases of the matrix elements can occur. \\

The search for \nnbb \, evidence spreads over many isotopes and different detection techniques.  The allowed \tnbb \, has been observed with several nuclei (e.g. $^{100}Mo, ^{82}Se, ^{48}K, ^{76}Ge,^{116}Cd, ^{136}Xe$) , which are naturally all potential candidates for \nnbb. Depending on the choice of isotope, the experimental searches differ in background performances, energy resolution, detection efficiencies as well as technical feasibility and available amounts of isotopes. Suppression of background has high priority, as in a background free measurement the sensitivity to  \mee \, scales with the square root of exposure instead of fourth root within a background limited search \cite{elliott}.

\subsection{Status}
The middle panel of table \ref{example_table_2col} shows published results on \mee. The quoted range in brackets is due to the uncertainty of the theoretically calculated matrix elements. Depending on the choice of \Mnonue \, sub-eV sensitivities are reached. There is even a claim for evidence of \nnbb , which has been critiqued by several authors \cite{aalseth} and is subject of verification by upcoming experiments, especially GERDA and Majorana  using the same detection technique. The upper limits from Cuoricino -depending on the assumption of matrix elements- start to exclude the claimed evidence.
Whereas the quoted experiments IGEX and HDM are finished, the experiments Cuoricino and NEMO3 progress to take data while being at the same time testbeds for next generation experiments. For a complete overview of published \mee \ results the reader is advised to the Double Beta Decay listings of the {\rm Particle Data Group}\cite{pdg}.

\subsection{Perspectives}
Substantial efforts are undertaken in \nnbb \, searches to access the \mee=50 meV region to distinguish between normal and inverted mass hierachy in the neutrino sector. This sensitivity calls for progress in background reduction as well as handling of target masses in the ton range. The Ge-experiments GERDA and MAJORANA focus in first phases on the demonstration of background suppression down to a level of $b<10^{-3}cts/(kg\cdot y \cdot  keV)$\cite{majorana2}. In these phases the experiments have already the sensitivity to fully explore the claimed evidence by \cite{klapdor2} . The question if the sensitivity is high enough to distinguish already at that stage between degenerated models will remain on the choice of matrix elements. GERDA is expected to be commissioned in 2009 and expects 1 year of data taking. The CUORE experiment aims to start measurement in 2011 and anticipates a required measuring time of 5 years to reach the 50 m\eV\ sensitivity. In their final phases GERDA, MAJORANA and EXO aim for detector masses in the ton-range, than being sensitive to the mass range of 10-50 meV.

\section{\label{sec_singleb} Limits and Sensitivity from single beta decay}
\subsection{Method}
The  energy spectrum of $\beta$ decay electrons
provides a sensitive direct and model independent search for the absolute electron
neutrino mass \cite{weinheimer}. The electron energy spectrum for \bdec\ for
a neutrino with mass $m_\nu$ is given by
\begin{eqnarray*}
{dN \over dE} & = & C \times F(Z,E) p E(E_0-E)   \\
              &  & \times [(E_0-E)^2-m_\nu^2]^{1
\over 2}\Theta (E_0-E-m_\nu) \label{mother}
\end{eqnarray*}
where $E$ denotes the electron energy, $p$ is the electron
momentum,  $E_0$ corresponds to the total decay energy, $F(Z,E)$
is the Fermi function, taking into account the Coulomb interaction
of the outgoing electron in the final state, the step function
$\Theta (E_0-E-m_\nu)$ ensures energy conservation, and $C$ is
given by
\begin{equation}
 C=G_F^2 {m_e^5  \over 2 \pi^3} \cos^2 \theta_C |M|^2 ~.
\label{re}
\end{equation}
 Here, $G_F$ is the Fermi constant, $\theta_C$ is the
Cabibbo angle, $m_e$ the mass of the electron and $M$ is the
nuclear matrix element. As both  $M$ and $F(Z,E)$ are independent
of $m_\nu$, the dependence of the spectral shape on $m_\nu$ is
given by the phase space factor only. \\
 A high precision measurement of the
electron energy is needed to resolve the count rate suppression
and spectrum distortion due to a massive \nueb, which are most
significant near the endpoint energy E$_0$.
Due to phase space arguments isotopes with low Q-value are favourable.

\subsection{Status}
The almost ideal features of tritium as a $\beta$ emitter have
been the reason for a long series of tritium \bdec\ experiments
\protect\cite{LANL,Zuerich,Tokyo,Bejing,LLNL}. The error bars on
the observable $\me^2$ \ of the various tritium \bdec\ experiments
over the last decade have decreased by nearly two orders of
magnitude. Equally important is the fact that the problem of
negative values for \mtwo\ of the early nineties has disappeared
due to better understanding of systematics and improvements in the
experimental setups. The last
experiments were performed by the Mainz\protect\cite{mainz}
and Troitsk\protect\cite{troitsk} group. The high sensitivity
of the Troitsk and the Mainz neutrino mass experiments is due to a
 type of spectrometers, so-called MAC-E-Filters
(\underline{M}agnetic \underline{A}diabatic
\underline{C}ollimation combined with an
  \underline{E}lectrostatic Filter)\protect\cite{beamson}. It combines high luminosity
and low background with a high energy resolution, both essential
to measure the neutrino mass from the endpoint region of a \bdec\
spectrum. The current results of both experiment yield upper
limits of   $\mnue \leq 2.1 \, \eV (\mbox{Troitsk})$ \cite{troitsk}, and
 $\mnue \leq 2.3 \, \eV (\mbox{Mainz})$ \cite{mainz}. \\

 An alternative approach is the use of calorimetric bolometers, with the absorber material being at the same time detector and source. Here, Rhenium ($^{187}$Re) the beta emitter with the lowest Q-value ($Q= 2.5\,$keV) can be used. With MIBETA and MANU two different Re bolometer techniques have been operated in the past, demonstrating the principle of operation yielding neutrino mass limits of   $\mnue \leq 15 \, \eV (\mbox{Mibeta})$ \cite{mibeta}, and
 $\mnue \leq 26 \, \eV (\mbox{Manu})$ \cite{manu} with $95\%$\,C.L.

\subsection{Perspectives}
The Karlsruhe Tritium Neutrino experiment (KATRIN)  with a large MAC-E filter  (10m diameter, 23m length, $\Delta$E =0.93\,\eV at the tritium endpoint energy) is under construction
to achieve a sensitivity of $\me < 0.2\, \eV \,(90\% C.L.)$ with statistical and systematic uncertainties contributing about equally. The experiment is expected to start in 2011.
Due to the exposed significance of a model-independent neutrino mass measurement a second approach with bolometric measurements  is proposed to follow KATRIN. The MARE II \cite{mare2,mare22} experiment would measure the  beta decay ($^{187}$Re) in a completely different approach from the point of view of experimental systematic uncertainties. The start of the experiment is envisaged at the end of next decade.

\section{\label{sec comparison} Comparison of Methods}

The highest sensitivity on the mass scale of neutrinos comes from cosmological observations. It has to be pointed out that the quoted limits on \Mnue \ are only valid within the used cosmological model and also depend on the priors used for the parameters. Additionally, there is also model-dependance due to astrophysical uncertainties e.g. the bias between dark matter and galaxies \cite{lahav_sys}.

Input from laboratory measurements will help to improve the systematics of cosmological analysis. A showcase for this is the correlation of the equation of state of dark energy $"w"$ with \Mnue \ in a flat  $\Lambda$ cold dark matter standard model. Reference \cite{elgaroy_k} shows that the model-independant measurements of \me \ by the KATRIN experiment help to break this degeneracy and improve significantly the data fits on $w$.
Also it as been shown (e.g. \cite{eitel},\cite{hannested_comb}) that the combining of laboratory measurements with cosmological observations improves siginificantly the sensitivity on \Mnue \ and help to constrain cosmological models. \\

From the laboratory measurements the masses from \nnbb \, show a higher sensitivity compared to the single beta decay method. Nevertheless, the model dependance arises from the fact that \nnbb \ can only occur if neutrinos are Majorana particles. On the other hand, examining the experimental possibilities of future experiments it looks like as  the \nnbb \ searches are the only way to explore neutrino masses below 100\, m\eV \, under laboratory conditions. As the uncertainty on the nuclear natrix element \Mnonue \, is a severe drawback for the \mee \ measurement, efforts are undertaken to improve the calculations. For example, in reference \cite{rodin} it is claimed that the uncertainty can be reduced to $30\%$ when the matrix elements are computed within a continuum QRPA ansatz.\\

Model-independant limits arise exclusively from single beta decay analysis. Theses limits can be regarded as conservative as it always holds that $\me\geq \mee$.

\section{Conclusions}
Several next generation experiments are under way, all with the sensitivity to answer within the next years if neutrinos are degenerated and if the neutrino mass is a crucial parameter for cosmological questions. As the methods are complimentary to each other in the sense that they measure different superpositions of the mass eigenstates a reliable answer to these fundamental questions can be expected. The most clean answer to that will come from the KATRIN experiment. For exploring mass regions below 100 meV the model-dependant approaches of cosmology and \nnbb \ have to be applied, at least on the time scale of the next decade.

\bigskip 


\end{document}